\documentclass[letterpaper, 10 pt, conference]{ieeeconf}  

\IEEEoverridecommandlockouts                              

\usepackage{graphics} 
\usepackage{epsfig} 
\usepackage{mathptmx} 
\usepackage{times} 
\usepackage{amsmath} 
\usepackage{amssymb}  
\usepackage{graphicx}
\usepackage{hyperref}
\usepackage{graphics,graphicx,amssymb,amsmath,verbatim}
\usepackage{mathrsfs}
\usepackage{amsfonts}
\usepackage{epstopdf}
\usepackage{xcolor}
\usepackage{subfigure}
\usepackage{hyperref}
\usepackage{booktabs}

\usepackage{algorithm}
\usepackage{algorithmic}
\usepackage{subfigure}
\usepackage{makecell}
\usepackage{diagbox}
\usepackage{multirow} 

\newenvironment{proof of Proposition }[1][Proof of Proposition]{\noindent\textit{#1.} }{\ \rule{0.5em}{0.5em}}
\newenvironment{proof of Theorem 2}[1][Proof of Theorem 2]{\noindent\textit{#1.} }{\ \rule{0.5em}{0.5em}}
\newenvironment{proof of Theorem 3}[1][Proof of Theorem 3]{\noindent\textit{#1.} }{\ \rule{0.5em}{0.5em}}
\newenvironment{proof of Theorem 4}[1][Proof of
Theorem 4]{\noindent\textit{#1.} }{\ \rule{0.5em}{0.5em}}
\newenvironment{proof of Theorem 5}[1][Proof of Theorem 5]{\noindent\textit{#1.} }{\ \rule{0.5em}{0.5em}}

\makeatletter
\def\footnoterule{\kern-3\p@
  \hrule \@width 2in \kern 2.6\p@}
\makeatother

\title{\LARGE \bf
Fast-SDE: Efficient Single-Microphone Sound Source Distance Estimation in Reverberant Environments
}

\author{\textit{Jiang Wang, Runwu Shi, Yaozhong Kang, Benjamin Yen, Takeshi Ashizawa, and Kazuhiro Nakadai}\\ 
\text{  }\\
Department of Systems and Control Engineering, Institute of Science Tokyo\\
{\small \texttt{\{wangjiang,shirunwu,kangyaozhong,benjamin,ashizawa,nakadai\}@ra.sc.eng.isct.ac.jp}}
\thanks{This work was supported by JST BOOST, Grant No. JPMJBS2430.}
	}
\begin{document}

\maketitle
\thispagestyle{empty}
\pagestyle{empty}

\begin{abstract}
Sound source distance estimation (SDE) is a critical capability in human–robot interaction. An inappropriate interaction distance not only reduces the reliability of speech acquisition and understanding, but also compromises the naturalness and comfort of the interaction. Most existing SDE methods rely on microphone arrays, however, multi-microphone systems typically require careful hardware synchronization, geometric calibration, and additional space and computational resources, which limits applicability to size-constrained and computability-limited embodied platforms. To alleviate these issues, we propose Fast-SDE, a lightweight single-microphone SDE framework that is suited for deployment on robot platforms with limited computational resources and strict size constraints. Specifically, Fast-SDE employs a subband-based backbone that decomposes the frequency axis into multiple subbands, rather than processing the entire spectrum with a wide full-band backbone. A shared subband encoder then maps each subband to a compact latent representation and learns the relationship between acoustic structure and time–frequency patterns. Finally, a lightweight regression head converts the fused subband representations into the estimated distance. Extensive simulation and real-world experiments demonstrate the merits of the proposed method. To benefit the broader research community, we have open-sourced our code at https://github.com/JiangWAV/FAST-SDE.
\end{abstract}

\section{Introduction}
Sound source distance estimation (SDE) aims to estimate the distance between a receiver and a sound source \cite{Neri24}. Estimating source distance provides critical spatial information for a range of tasks in human–robot interaction \cite{HRI23}, including sound source localization \cite{Wang25iros,AnTRO,Fu2024}, speech enhancement and separation \cite{patterson2022distance,shiaaai,shierosip}, and active robotic exploration \cite{Younes2023}, thereby improving interaction quality and feedback. Most existing studies on sound source distance estimation rely on microphone arrays. By exploiting inter-channel cues such as phase difference (IPD) and intensity differences (IID)\cite{Rascon2017}, these systems can achieve accurate spatial perception. 

\begin{figure}[t]
\centering 
{\includegraphics[width=0.8\columnwidth]{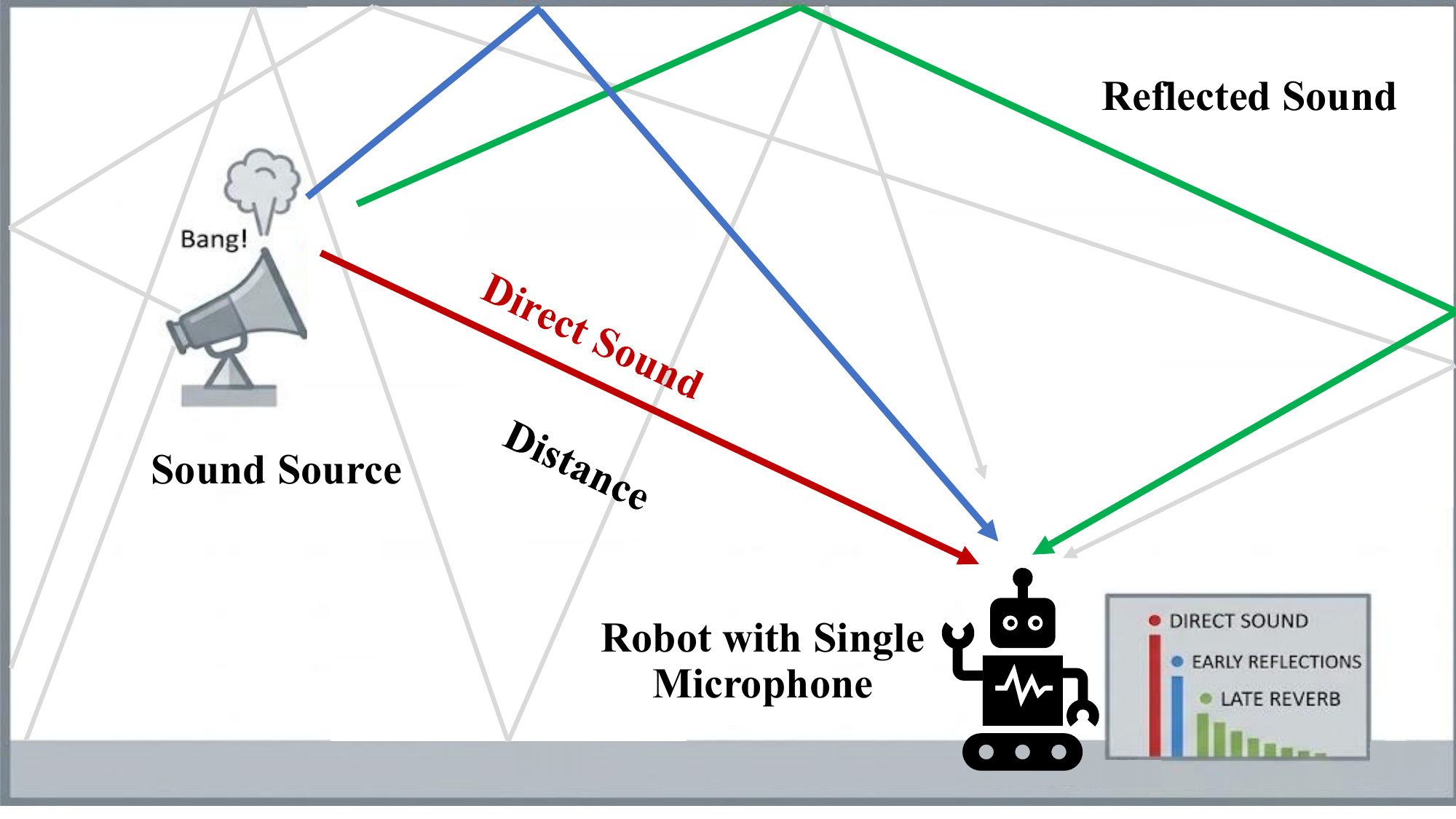}
}
\caption{Problem setup of sound source distance estimation using a single microphone in reverberant environments.}
\label{ssl} 
\vspace{-1.5em}
\end{figure}

However, multi-microphone systems usually require careful hardware synchronization, geometric calibration, and additional computational and communication resources \cite{Kong2021,Li24,Jiang24}. These requirements limit their applicability to low-cost, size-constrained, and energy-limited platforms. In contrast, a single microphone offers an appealing alternative due to its lower hardware complexity, lower power consumption, and easier deployment. Despite its practical importance, single microphone SDE remains highly challenging. Unlike microphone arrays, a single microphone cannot directly access explicit spatial cues such as the relative position of the microphones. Instead, the distance must be inferred indirectly from subtle acoustic patterns embedded in the received waveform or spectrogram \cite{Georganti11}. In real environments, however, these cues are strongly affected by room acoustics, source-content variability, background noise, and microphone characteristics. In particular, as shown in Fig. \ref{ssl}, reverberation can severely obscure the relationship between the observed signal and the source distance. Thus, monaural SDE remains an open and challenging problem.

Recently, learning-based methods have shown promising performance in monaural distance estimation by using neural networks to learn implicit distance-related representations from large-scale data \cite{Neri24,Wang25iros}. Nevertheless, these approaches rely on relatively heavy models and high computational cost, which limits their suitability for real-time applications and resource-constrained platforms. For practical deployment, especially on robot platforms, an SDE model must be not only accurate but also lightweight and fast.

In this paper, we propose a computationally efficient monaural SDE framework for reverberant environments. Specifically, we introduce a subband-based backbone that decomposes the frequency axis into multiple non-overlapping subbands instead of processing the full spectrum with a wide backbone. This design preserves distance-discriminative acoustic information while reducing model complexity and inference latency. A shared subband encoder then maps each subband to a compact latent representation and learns the correlation between acoustic structure and time–frequency patterns at low computational cost. Finally, a lightweight regression head converts the fused subband representation into the estimated distance. Owing to this design, Fast-SDE is well suited for deployment on robot platforms with limited memory and computing resources. Experiments in both simulated and real-world environments show that the proposed method achieves competitive distance estimation accuracy with fewer parameters and lower runtime cost than existing approaches.

\section{Related Work}

In this section, we discuss existing methods that infer spatial information of a sound source from monaural audio.

\subsection{Monaural DOA Estimation}
Some monaural direction of arrival (DOA) estimation methods exploits artificial pinna-like structures, which act as acoustic filters and shape the transfer function from the sound source to the microphone in a direction-dependent manner. Because this filtering effect varies with the source direction, it provides discriminative spectral cues for DOA estimation. For example, \cite{Andrew09} trained a hidden Markov model to represent the direction-dependent transfer functions induced by a pinna, thereby enabling direction estimation. \cite{Youssef24} extracted the second-order derivatives of spectra corresponding to different source directions and used them as inputs to a multilayer perceptron for DOA estimation. 
When no artificial structure is available, monaural DOA estimation instead rely on room-related transfer characteristics. \cite{Takashima13} proposed a cepstral feature weighting method based on multiple kernel learning, which selectively reweights cepstral dimensions associated with the sound transfer function. \cite{Guo22} introduced a composite reverberant speech model and a directly trained reverberant speech model to estimate room acoustic transfer functions and speaker direction from reverberant speech.

\subsection{Monaural Sound Source Distance Estimation}
Compared with DOA estimation, sound source distance estimation is generally considered a more challenging task \cite{Neri24}. Early study \cite{Georganti11} mainly relied on low-level acoustic features, such as linear predictive coding, skewness, and spectral kurtosis, to classify the distance of a speaker, but it was unable to estimate a precise distance. \cite{Vetterli14} associated recorded reverberation with the reflecting walls that generated it and formulated a least-squares problem to estimate the source-to-microphone distance. However, their method is not suitable for sustained source signals. More recently, \cite{Neri24} employed a convolutional recurrent neural network with an attention module to infer the source-to-microphone distance from reverberant audio. Similarly, \cite{Wang25iros} used an attention-based architecture to estimate the source-to-microphone distance in reverberant environments and further combined it with an extended Kalman filter to infer the sound source position. However, these attention-based methods usually incur relatively high computational cost and large model size. Such limitations reduce their suitability for real-time applications and resource-constrained platforms. To this end, we propose a more lightweight monaural SDE framework that is better suited for deployment on hardware platforms with limited computational resources, such as embedded devices, while still achieving competitive distance estimation accuracy.

\section{Proposed method}
\subsection{Setup and Problem Statement}
We consider a single microphone that receives reverberant acoustic signals emitted by a stationary sound source in an enclosed room. Denote $s[n]$ as the clean source signal, where $n$ is the discrete-time sample index, and let $h\left[n;p,m,\varepsilon\right]$ be the room impulse response (RIR) between the source at position $p\in\mathbb{R}^{2}$ and the microphone at position $m\in\mathbb{R}^{2}$, where $\varepsilon$ summarizes latent environmental factors (e.g., room geometry, surface materials). Therefore, the received reverberant signal is 
\begin{equation}
x[n]=h\left[n;p,m,\varepsilon\right]\ast s[n]+v[n],
\end{equation}
where $\ast$ denotes convolution and $v[n]$ is additive noise.

In reverberant environments, the RIR can be conceptually decomposed into a direct-path component $h_{dir}\left[n;p,m,\varepsilon\right]$ and a reverberant component $h_{rev}\left[n;p,m,\varepsilon\right]$ \cite{Samarasinghe} as follows:
\begin{equation}
h\left[n;p,m,\varepsilon\right]=\ensuremath{h_{dir}\left[n;p,m,\varepsilon\right]}+\ensuremath{h_{rev}\left[n;p,m,\varepsilon\right].}
\end{equation}
By applying the short-time Fourier transform (STFT), the monaural signal can be written as
\begin{equation}
X(i,f)=\sum_{k=0}^{K-1}\left(H_{dir}\left(k,f\right)+H_{rev}\left(k,f\right)\right)S\left(i-k,f\right)+V\left(i,f\right),\label{STFT}
\end{equation}
where $i\in[1,T]$ and $f\in[1,F]$ denote the time-frame index and frequency-bin index \cite{wang2023tf}, respectively. $X(i,f)$, $S(i,f)$, and $V(i,f)$ are the STFT coefficients of the microphone signal, source signal, and additive noise, respectively, while $H_{dir}\left(k,f\right)$ and $H_{rev}\left(k,f\right)$ denote the convolutive transfer functions associated with the direct-path and reverberant components in the STFT domain.


In contrast to multi-microphone systems \cite{Rascon2017,IPD}, where microphone pairs provide robust spatial cues, the single-channel observation model in (\ref{STFT}) must infer the target distance indirectly from reverberation-dependent acoustic patterns. Consequently, the problem considered in this work is to estimate the source-to-microphone Euclidean distance $\hat{d}(p,m)$ from distance-related acoustic cues embedded in the single-channel audio $x[n]$.

\begin{figure*}[t]
\centering 
\includegraphics[width=0.80\textwidth]{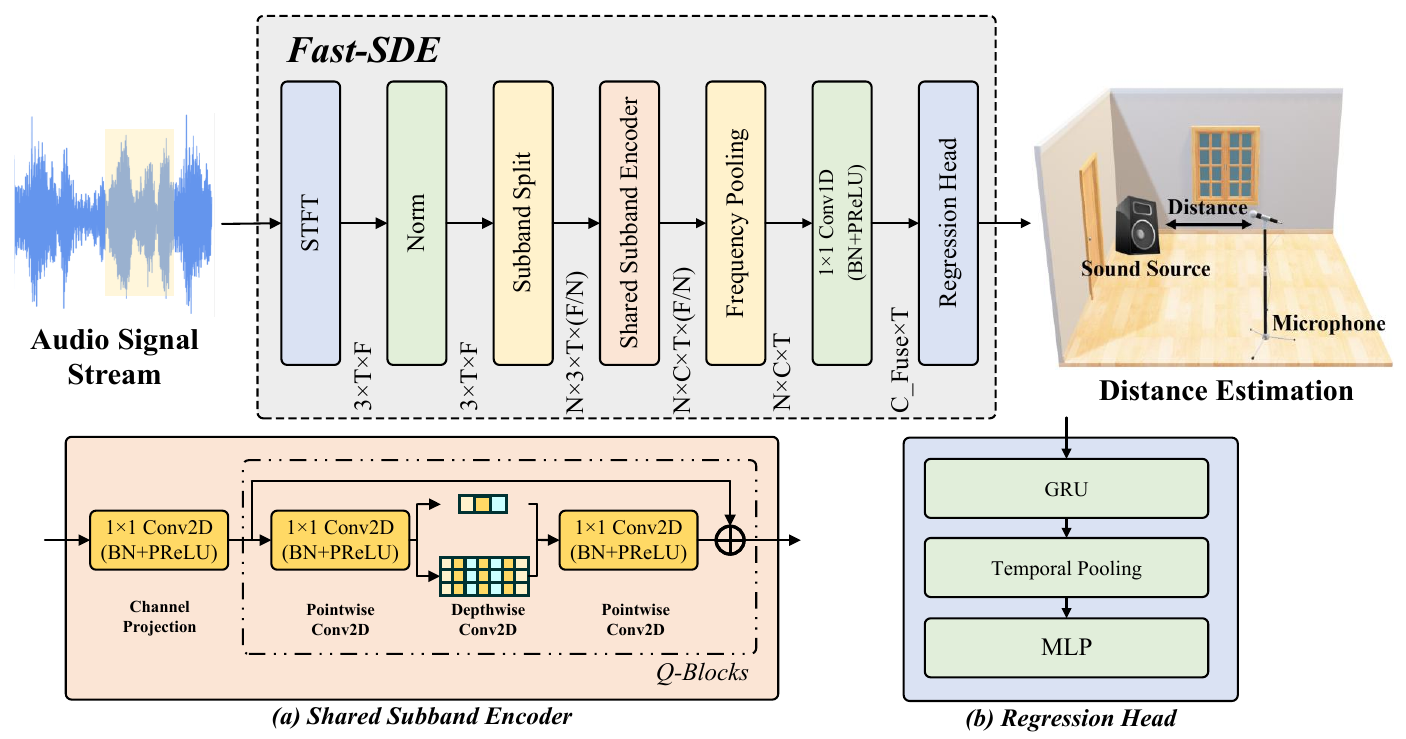}
\caption{Structure of Fast-SDE (a) Shared Subband Encoder (b) Regression Head. }
\label{network} 
\vspace{-1em}
\end{figure*}

\subsection{Learning Target}
The signal received by a single microphone is shaped by the acoustic propagation channel, which jointly affects the amplitude and phase of the recorded waveform. Specifically, the log-magnitude STFT captures energy variations introduced during propagation, such as distance-dependent attenuation and frequency-dependent absorption by surfaces. In contrast, the phase spectrum provides cues about the underlying acoustic path structure, e.g., due to multipath propagation, the phase patterns vary with the relative strengths and delays of the direct-path and reflected components. Moreover, to obtain a smoother representation that is more suitable for neural networks, we encode phase using its sine and cosine maps, which preserve phase information while avoiding discontinuities caused by $2\pi-wrapping$.
Therefore, we construct a three-channel time–frequency feature input
\begin{equation}
\mathbf{X}_{in}=\left[\log\left(\left|X\right|^{2}\right),\cos\left(\angle X\right),\sin\left(\angle X\right)\right]\in \mathbb{R}^{3 \times T \times F}
\end{equation}
and learn a mapping from $\mathbf{X}_{in}$ to the source–microphone distance.

\subsection{Network Architecture}
As illustrated in Fig. \ref{network}, Fast-SDE is a lightweight network for estimating source-to-microphone distance from a short single channel audio stream. Overall, Fast-SDE consists of a subband splitting module, a shared subband encoder, a frequency pooling and fusion stage, and a regression head. Given an input waveform segment, the model first computes its STFT and constructs a compact time–frequency feature representation by stacking the logarithmic magnitude spectrum and the sine/cosine phase maps.  The resulting tensor $\mathbf{X}_{in}$ is normalized per-segment and then fed into the subband splitting module. The subbands are processed by a shared encoder built with lightweight pointwise and depthwise 2D convolutions to extract compact local reverberation cues. Subsequently, these subband representations are fused along the temporal axis, and the final source–microphone distance is predicted by a lightweight regression head composed of gated recurrent unit (GRU)-based temporal modeling, temporal pooling, and a multi-layer perceptron (MLP).

\textbf{Subband-Based Backbone:} 
The backbone of Fast-SDE operates on frequency subbands rather than on the full spectrum with a wide feature extractor. Specifically, the input feature map is decomposed into $N$ non-overlapping subbands, where each subband corresponds to a local frequency region. This design reduces computational cost and makes the model more suitable for lightweight deployment. At the same time, it enables the network to capture band-specific structures more effectively, since low- and high-frequency components often exhibit different statistical and perceptual characteristics \cite{Luo24}.  By restricting each branch to a local frequency region, the model can focus on informative local patterns while reducing interference from irrelevant distant frequency regions.  All subbands are processed by the same shared subband encoder, meaning that a single set of parameters is reused across all subbands.  In implementation, the computation is vectorized by merging the subband index into the batch dimension, which allows all subbands to be processed efficiently in parallel.

\textbf{Shared Subband Encoder:} 
The shared subband encoder maps each subband to a compact latent representation while maintaining low computational cost.  It begins with a $1 \times 1$ convolution followed by batch normalization (BN) and parametric rectified linear unit (PReLU), which projects the three feature channels input into $C$ intermediate feature channels. The projected representation is then refined by $Q$ lightweight residual blocks. Each block adopts a pointwise–depthwise–pointwise 2D convolution design. First, a $1 \times 1$ convolution mixes channel information. Next, depthwise 2D convolutions extract local time–frequency patterns within each subband. In particular, two depthwise convolution branches with different receptive fields are used to capture complementary temporal and spectral structures at different scales. Finally, another $1 \times 1$ convolution fuses the responses back into the original channel dimension. Residual connections preserve low-level information and stabilize optimization when multiple blocks are stacked. Since the encoder operates independently within each subband, it produces an output tensor of shape $\mathbb{R}^{N \times C \times T \times (F/N)}$.

\textbf{Frequency Pooling and Subband Fusion:} 
To obtain a compact descriptor for each subband, Fast-SDE applies frequency pooling along the within-subband frequency axis. The pooled subband descriptors are then concatenated along the channel dimension and passed through a $1 \times 1$ Conv1D layer with BN and PReLU to generate a unified temporal feature sequence. This fusion strategy provides an efficient way to integrate information across subbands without introducing computationally expensive global operations over the entire $T\times F$ grid. As a result, the model preserves the advantages of subband-wise processing while still allowing information exchange across frequency regions at a later stage.

\textbf{Regression Head:} 
As shown in Fig. \ref{network}(b), the regression head converts the fused temporal sequence into the final distance estimate.  A lightweight GRU first models the temporal evolution of reverberation cues across frames. The GRU outputs are then aggregated by temporal pooling to form a fixed-dimensional segment-level representation, which is finally mapped to a scalar distance by a small MLP.  This design keeps the temporal modeling module compact while still enabling the network to exploit both short-term fluctuations and long-term decay patterns in the reverberant observation.  Nevertheless, since GRU-based temporal modeling remains relatively expensive, we further simplify the regression head in the UltraFast-SDE variant by replacing the GRU with a purely feed-forward design consisting of temporal pooling followed by an MLP, which yields a faster and more lightweight architecture.

\subsection{Model Implementation and Training}
All input audio signals were resampled to 16 kHz. We computed the STFT using a window length of 512 samples (32 ms) a hop size of 128 samples (8 ms). The main model hyperparameters are summarized in Table \ref{tab:model_parameters}. To accommodate different computational budgets, we implemented two model variants, namely Fast-SDE and UltraFast-SDE. For Fast-SDE, the frequency axis is divided into 8 subbands. The shared subband encoder contains 3 residual blocks, each using 64 intermediate feature channels, and the temporal modeling module is implemented with a unidirectional GRU whose hidden size is 96. Then, UltraFast-SDE is designed for scenarios with more limited computational resources. In this variant, the frequency axis is divided into 6 subbands, and the shared subband encoder contains 2 residual blocks with 48 intermediate feature channels. Moreover, the GRU module is removed, and the regression head is simplified to temporal pooling followed by an MLP, resulting in a more lightweight and efficient architecture.

During training, we used the Adam optimizer with an initial learning rate of 0.001 and a decay factor of 0.8. The batch size was set to 84, and the model was trained by minimizing the mean squared error (MSE) between the predicted and ground-truth distances.

\begin{table}[t]
  \centering
  \caption{Model Implementation Parameters.}
  \scalebox{0.86}{
  \begin{tabular}{ccccc}
    \toprule
    Model & Hyperparameters & Value & Hyperparameters & Value \\
    \midrule
    \multirow{3}{*}{Fast-SDE}&Frequency Bins & 256 & Subbands (N)     & 8 \\ &Residual Blocks (Q)& 3 &Encoder Channels (C)    & 24   \\   
    &C\_Fuse & 64 & GRU hidden state       & 96  \\
    \midrule
    \multirow{3}{*}{UltraFast-SDE}&Frequency Bins & 256 &Subbands (N)  & 6\\  &Residual Blocks (Q)& 2 &Encoder Channels (C)    & 16 \\
    &C\_Fuse & 48      & GRU hidden state       & -- \\
    \bottomrule
  \end{tabular}}
  \label{tab:model_parameters}
  \vspace{-1.0em}
\end{table}

\begin{table}[t]
    \begin{center}
    \renewcommand\arraystretch{1.2}
    \caption{\label{tab:dataset}Simulations Dataset Parameters}
    \begin{tabular}{cccccc} 
    \toprule
    Dataset & Room & RT60 (s) & Mic num & Src per mic & Total\\
    \midrule
    Group\_1 & 1  & 0.6  & 900 & 50 & 45,000 
    \\
    Group\_2 & 10 & 0.4 - 0.75 & 100 & 50 & 50,000 
    \\
    Group\_3 & 100 & 0.4 - 0.75 & 100 & 50 & 500,000  
    \\
\toprule
\end{tabular}
\end{center}
\vspace{-1.5em}
\end{table}

\section{Simulation Experiments}
\begin{table*}[t]
    \begin{center}
    \renewcommand\arraystretch{1.2}
    \caption{\label{tab:simulation}Comparasion of Sound Source Distance Estimation With Other Method (Bold Means Better)}
    \centering 
    \begin{tabular}{ccccccccc}
    \toprule [1pt]
    \multicolumn{1}{c}{\multirow{3}{0.5cm}{Models}} 
    & \multicolumn{3}{c}{MAE (m)} 
    & \multicolumn{1}{c}{\multirow{3}{0.5cm}{Params}}
    & \multicolumn{1}{c}{\multirow{3}{0.5cm}{FLOPs}}
    & \multicolumn{3}{c}{Inference Latency (ms)} \\
    \cmidrule(lr){2-4}
    \cmidrule(lr){7-9}
    &  \makecell[c]{ Group\_1} & \makecell[c]{Group\_2} & \makecell[c]{Group\_3 }&  &  & \makecell[c]{GPU\\ (V100)}& \makecell[c]{CPU\\ (R5 4500U)}& \makecell[c]{Micro-controller\\ (ESP32-S3-Zero)} \\
    \midrule [1pt]
    Fast-SDE (Our) & \textbf{0.13} & 0.19 & \textbf{0.23} & 75.8K & 121.0M& 2.09 & 9.84 & 2837\\
    UltraFast-SDE (Our) & 0.18 & 0.26 & 0.26 & \textbf{9.6K}& 50.9M& \textbf{1.61} & \textbf{5.23}  & \textbf{1018} \\
    SELDNet\cite{Neri24} & \textbf{0.13} & \textbf{0.18} & 0.24 & 649.4K& 187.2M & 2.16 & 18.19& -- \\
    Attention\cite{Wang25iros} & 0.19 & 0.29 & 0.31 & 42.9K& \textbf{34.8M}& 13.61& 30.88& 2452  \\
\toprule
\end{tabular}
\end{center}
\vspace{-1.5em}
\end{table*}

\subsection{Datasets}
To simulate reverberation in indoor environments, we used FRAM-RIR \cite{luo2024fast} to generate three datasets, namely Group\_1, Group\_2, and Group\_3. Each sample was annotated with its corresponding source-to-microphone distance. We then convolved a 0.2 s chirp signal with the generated room impulse response (RIR) to simulate the audio recorded by the microphone \cite{liu2025}.

The three datasets differ in the number of rooms used to generate the RIRs, namely 1, 10, and 100 rooms for Group\_1, Group\_2, and Group\_3, respectively. This design allows us to evaluate the model under room configurations of increasing diversity and, consequently, to examine its generalization ability across unseen acoustic conditions. Table \ref{tab:dataset} summarizes the dataset configurations, where Mic num denotes the number of microphones in each room, Src per mic denotes the number of source positions associated with each microphone, and reverberation time (RT60) is used to simulate various room acoustic environments.

For each dataset, we collected samples from all available rooms using different microphone and source locations, and then split the data into training, validation, and test sets with a ratio of 0.8:0.05:0.15. The room dimensions were randomly sampled within the range of 5.4m × 6.4m × 2.5m to 6.4m × 7.4m × 3.5m.

\subsection{Experimental Results}
To evaluate the proposed Fast-SDE framework, we conducted a series of simulation experiments. We first examined the model performance under different room configurations in order to assess its generalization capability. We then compared the proposed method with the state-of-the-art single microphone distance estimation approach, SeldNet \cite{Neri24}, as well as the Attention-based model in \cite{Wang25iros}. The comparison considered not only estimation accuracy, but also model complexity and inference latency on different hardware platforms, including an NVIDIA V100 GPU, an AMD R5 4500U CPU, and an ESP32-S3-Zero microcontroller.

\begin{figure}[t]
\centering 
{\includegraphics[width=0.9\columnwidth]{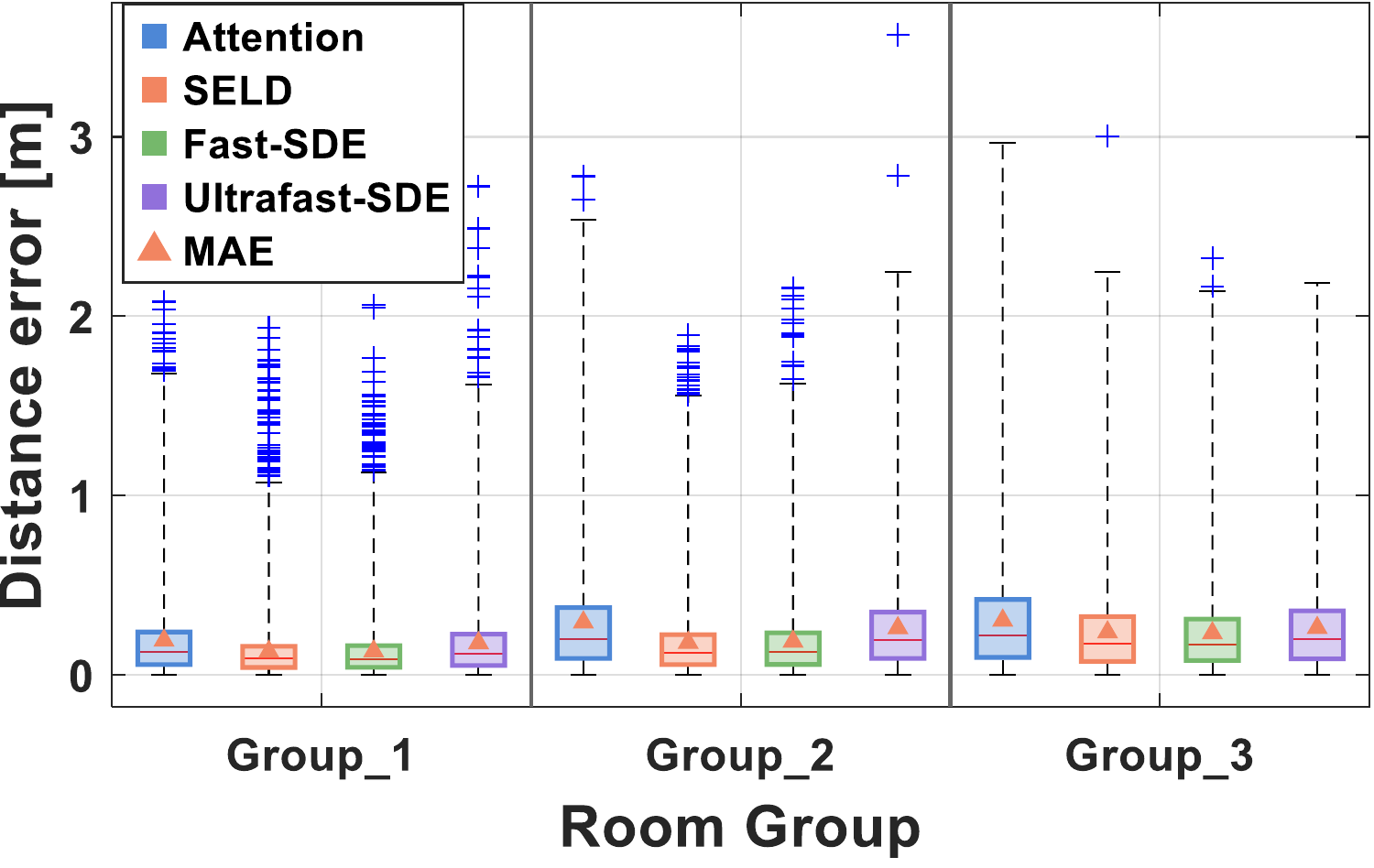}
}
\caption{Error distributions between the estimated sound source distance and true values for different datasets.}
\label{room} 
\vspace{-1.5em}
\end{figure}

We used the mean absolute error (MAE) as the metric for distance estimation accuracy. Table~\ref{tab:simulation} presents the quantitative comparison, while Fig.~\ref{room} shows the error distributions between the estimated and ground-truth distances on the four test sets. The results show that the estimation error increases as the diversity of acoustic environments increases, reflecting the greater difficulty of generalizing in more room configurations. Compared with the state-of-the-art methods for single microphone sound source distance estimation, the proposed approach achieves accurate distance estimation with fewer parameters, and lower inference latency. In particular, UltraFast-SDE remains deployable even on the widely used ESP32-S3 microcontroller with only 4 MB of flash memory and 2.5 MB of RAM. Despite these severe resource constraints, UltraFast-SDE still achieves one inference per second, demonstrating its potential for sound source distance estimation in highly resource-limited embedded systems.

\begin{figure}[t]
        \centering
	\begin{minipage}{0.7\linewidth}
		\centering
		\subfigure[]{\includegraphics[width=1\linewidth]{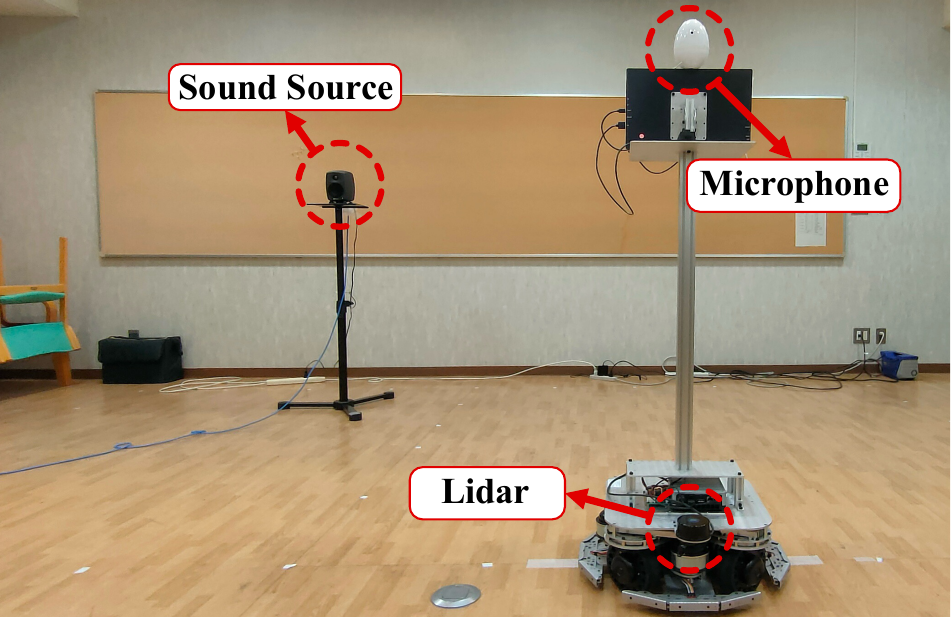}}
	\end{minipage}
	\centering
	\begin{minipage}{0.7\linewidth}
		\centering
		\subfigure[]{\includegraphics[width=1\linewidth]{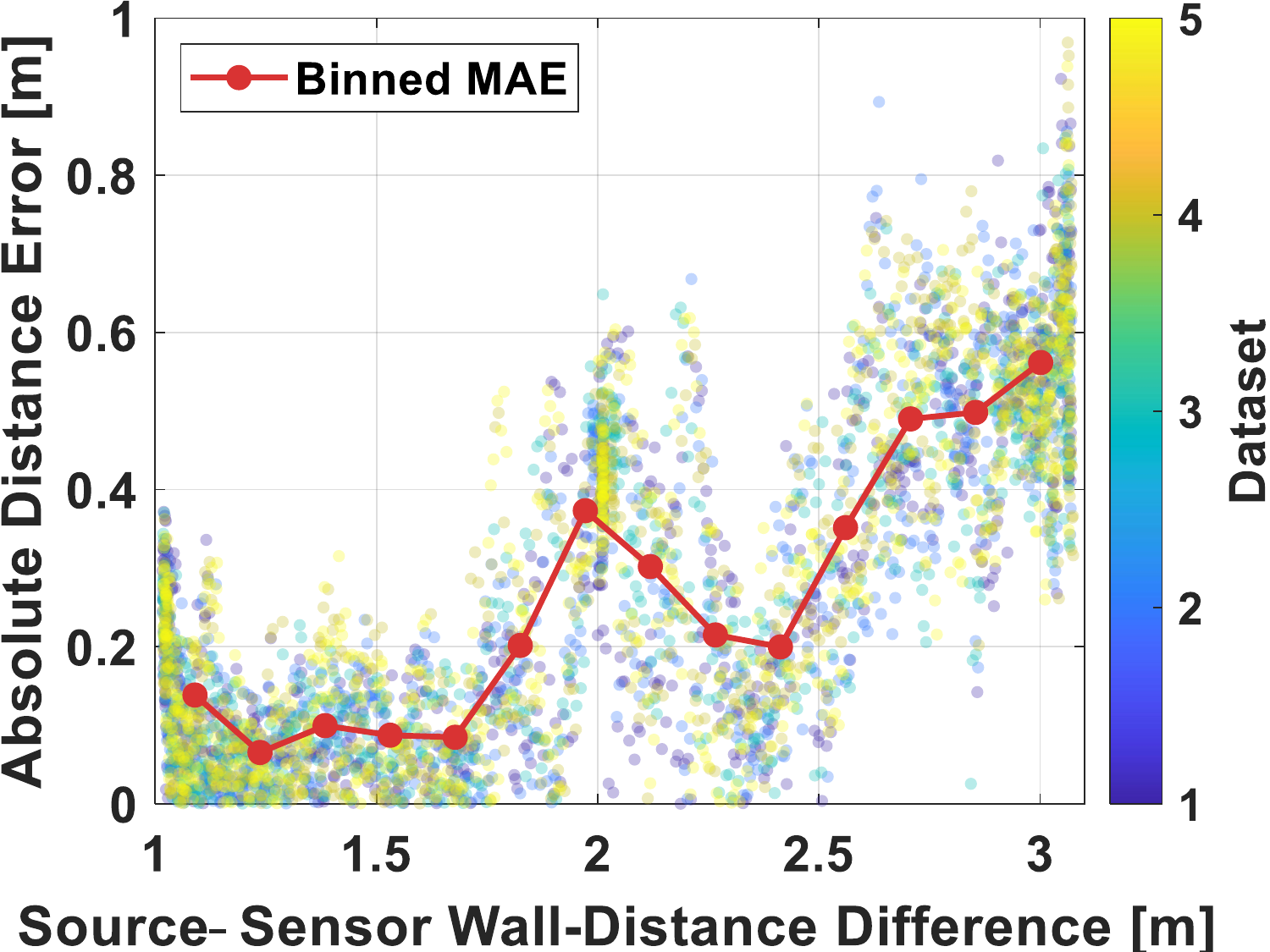}}
	\end{minipage}
	\begin{minipage}{0.8\linewidth}
		\centering
		\subfigure[]{\includegraphics[width=1\linewidth]{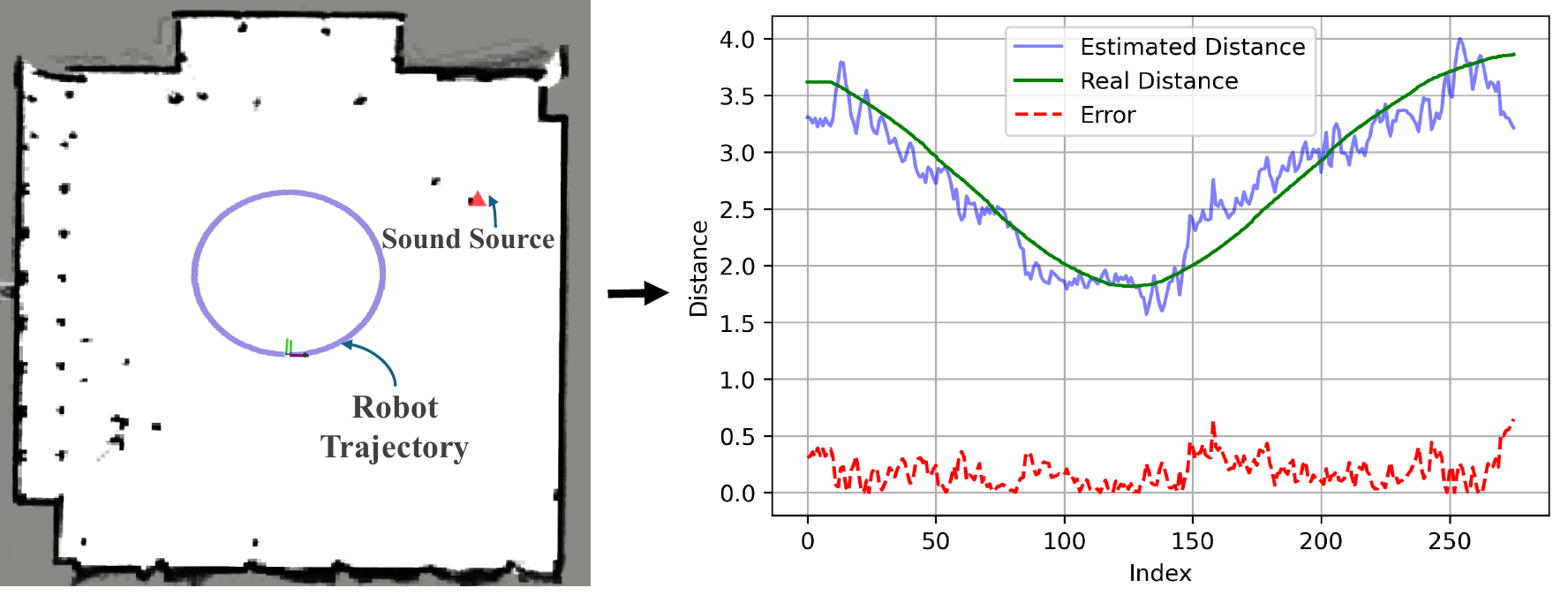}}
	\end{minipage}
\caption{Real-world sound source distance estimation using a single microphone: experiments setup and results (a) Experiment setup, (b) Correlation between SDE errors and the difference in distance from the source and microphone to the walls, (c) The SDE results at different time instance.}
    \label{fig:exp_scene}
\vspace{-1.5 em} 
\end{figure}

\section{Real World Experiments}
\subsection{Experimental Setup}
To further validate the proposed method, we conducted real-world sound source distance estimation experiments on a mobile robot platform. The experiments were performed using a Vstone 4WDS ROVERX40A robot equipped with an onboard computer powered by an AMD R5 3550H processor, a YDLiDAR TG30, and a TAMAGO-01 microphone, as shown in Fig. \ref{fig:exp_scene}(a). The microphone records audio at a sampling rate of 16 kHz with a 24-bit resolution while a loudspeaker continuously emits chirp signals. The LiDAR is used to estimate the robot pose, from which the ground-truth distance between the microphone and the loudspeaker is obtained.

To bridge the gap between simulation training and real-world deployment, we fine-tuned each model initially trained on the Group\_3 dataset. Specifically, we placed the sound source at five different positions in the target room \((5.9\,\mathrm{m} \times 6.9\,\mathrm{m} \times 2.9\,\mathrm{m})\), while the robot recorded audio along the trajectory shown in Fig. \ref{fig:exp_scene}(c). Each sampling was labeled with the corresponding source-to-microphone distance, and the model was then fine-tuned for 200 epochs. After fine-tuning, we placed the sound source in new positions to evaluate the model under real acoustic conditions.

\begin{table}[t]
    \begin{center}
    \renewcommand\arraystretch{1.2}
    \caption{\label{tab:4}SDE Errors at various sound source positions (Bold Means Better)(Unit: \text{m})}
    \centering 
    \scalebox{0.8}
    {\begin{tabular}{ccccccc}
    \toprule
    Models & Position A & Position B & Position C & Position D & MAE\\
    \midrule
    Fast-SDE (Our) & \textbf{0.20} &\textbf{0.19}  & 0.26 & \textbf{0.24} & \textbf{0.22} \\
    UltraFast-SDE (Our) & 0.26& 0.22 & 0.29 & 0.28 & 0.26\\
    SELDNet\cite{Neri24} & 0.26& 0.30 & \textbf{0.25} & 0.28 & 0.27\\
    Attention\cite{Wang25iros} & 0.24&  0.36&  0.31& 0.29 & 0.30 \\
\toprule
\end{tabular}}
\end{center}
\vspace{-2.3em}
\end{table}

\subsection{Sound Source Distance Estimation Results}
We carried out two groups of real-world experiments. In the first group, we investigated how the difference between the distance from the source and the microphone to the wall affects the accuracy of distance estimation. The source was fixed at \((2,3)\), while the robot moved along the trajectory shown in the Fig. \ref{fig:exp_scene}(c) and repeated five times, with each sample lasting 0.6 s. Then, the sound source distance estimation results are divided into 14 bins according to the source–microphone wall-distance difference. As shown in Fig. \ref{fig:exp_scene}(b) and (c), the right wall acted as the reflective surface, and the estimation error increased significantly as the wall-distance difference became larger. The correlation coefficient reached 0.77. These results indicate that the source–microphone wall-distance difference has a substantial effect on distance estimation performance.

In the second group of experiments, we evaluated the performance of different models in the real environment. We placed the sound source at four different positions, denoted as Position A--Position D, while the robot again followed the trajectory shown in  Fig. \ref{fig:exp_scene}(c) to collect data. We then computed the MAE for each dataset. The results in Table \ref{tab:4} show that the proposed Fast-SDE achieves the lowest estimation error among all compared methods in real-world.

\section{\label{CONCLUSION}CONCLUSION}
In this work, we investigated the task of estimating the source-to-microphone distance from a single microphone in reverberant environments and proposed a lightweight monaural SDE framework, termed Fast-SDE. The proposed framework enables real-time distance estimation from reverberant signals with limited computational resources and strict size constraints. Simulation results under different room configurations show that the estimation error increases as the diversity of acoustic environments grows, highlighting the challenge of generalization. Real-world experiments further reveal that the difference between the distances from the source and the microphone to the wall has a significant impact on distance estimation performance. In future work, we plan to extend the proposed framework to more challenging 3D scenarios involving multiple sound source types.

\end{document}